\documentclass[11pt,draftcls,onecolumn]{IEEEtran}

\usepackage{cite}
\usepackage[cmex10]{amsmath}
\usepackage{array}
\usepackage{fixltx2e}
\usepackage{amsfonts}
\usepackage{color}
\usepackage{amsmath}
\usepackage{amsthm}
\usepackage{amssymb}
\usepackage{t1enc}
\usepackage{alltt}
\usepackage{epsfig}
\usepackage{mathrsfs}
\usepackage{graphicx,ifthen}
\usepackage{multirow}
\usepackage{citesort}
\usepackage[scaled]{helvet}
\usepackage{booktabs}
\usepackage{wasysym}
\usepackage{arydshln}
\usepackage{enumerate}
\usepackage{bbm}

\allowdisplaybreaks[1]
\newcommand{\e}{{\rm e}}
\newcommand{\tr}{{\mathrm{tr}}}

\makeatletter

\newcommand{\Rmnum}[1]{\expandafter\@slowromancap\romannumeral #1@}
\newcommand{\ra}[1]{\renewcommand{\arraystretch}{#1}}

\newcommand{\dd}{\,\mathrm{d}}

\begin{document}
\title{From Random Matrix Theory to Coding Theory: Volume of a Metric Ball in Unitary Group}

\author{Lu~Wei,~Renaud-Alexandre~Pitaval,~Jukka~Corander,~and~Olav~Tirkkonen%
\thanks{L. Wei and J. Corander are with the Department of Mathematics and Statistics, University of Helsinki, Finland (e-mails: \{lu.wei, jukka.corander\}@helsinki.fi). R.-A. Pitaval is with the Department of Mathematics and Systems Analysis, Aalto University, Finland (e-mail: renaud-alexandre.pitaval@aalto.fi). O. Tirkkonen is with the Department of Communications and Networking, Aalto University, Finland (e-mail: olav.tirkkonen@aalto.fi).}%
\thanks{This work was presented in part at 2015 IEEE International Symposium on Information Theory.}}


\maketitle

\begin{abstract}
Volume estimates of metric balls in manifolds find diverse applications in information and coding theory. In this paper, some new results for the volume of a metric ball in unitary group are derived via various tools from random matrix theory. The first result is an integral representation of the exact volume, which involves a Toeplitz determinant of Bessel functions. The connection to matrix-variate hypergeometric functions and Szeg\H{o}'s strong limit theorem lead independently from the finite size formula to an asymptotic one. The convergence of the limiting formula is exceptionally fast due to an underlying mock-Gaussian behavior. The proposed volume estimate enables simple but accurate analytical evaluation of coding-theoretic bounds of unitary codes. In particular, the Gilbert-Varshamov lower bound and the Hamming upper bound on cardinality as well as the resulting bounds on code rate and minimum distance are derived. Moreover, bounds on the scaling law of code rate are found. Lastly, a closed-form bound on diversity sum relevant to unitary space-time codes is obtained, which was only computed numerically in literature.
\end{abstract}
\

\begin{IEEEkeywords}
Coding-theoretic bounds, random matrix theory, unitary group, volume of metric balls.
\end{IEEEkeywords}

\section{Introduction}
Determining the volume of metric balls in Riemannian manifold, in particular unitary group, is the key to understand several coding
and information theoretical quantities. Performance analysis of unitary space-time codes~\cite{2001Shokrollahi,2002Liang,2006bHan} requires the knowledge of volume in the unitary group~\cite{2006Han,2010Creignou}. For channel quantizations in precoded multi-antenna systems, the characterization of rate-distortion tradeoff is directly related to volume calculations in the manifold of interest~\cite{2008Dai,2013Krishnamachari}. Estimating fundamental coding bounds such as Gilbert-Varshamov and Hamming bounds relies on the volume of the corresponding metric ball~\cite{2002Barg,2005Henkel,2008Krishnamachari,2012Pitaval}.

Despite the need to accurately estimate the volume of metric balls in the unitary group, results in this direction are rather limited. Volume estimates have been derived in~\cite{2005Henkel,2008Krishnamachari,2012Pitaval} when the radius of metric ball is small. In this paper, we study the volume of a metric ball, valid for any radius, in the unitary group with chordal distance. The same problem was considered in~\cite{2006Han} in the study of diversity sum bounds of unitary space-time codes, where the authors relied on numerical integrations to evaluate the volume. The starting point of the current paper is that such a numerical step may not be necessary. Specifically, we show that the exact volume boils down to an integral involving a Toeplitz determinant with Bessel function entries. This representation gives rise to an explicit formula in the simplest case of a two-by-two unitary matrix. We then present a simple asymptotic volume formula, which is the main technical contribution of the paper. Two distinct paths that led to this asymptotically exact formula have been identified: one is based on the connection of the Toeplitz determinant to a hypergeometric function of matrix argument, the other directly invokes Szeg\H{o}'s strong limit theorem on Toeplitz determinants. These powerful tools of random matrix theory, albeit being subjects of intense studies for several decades, have not been fully utilized in the coding theory community. Surprisingly, the limiting volume formula is already quite accurate for dimension as small as three. The reason behind the rapid convergence is examined, where it is found that the asymptotic formula approaches its limit super-exponentially fast as dimension increases due to a discovered  mock-Gaussian property.

As an application, we study some basic coding-theoretic questions of unitary codes via the derived asymptotic volume formula. In particular, analytical expressions for the Gilbert-Varshamov and the Hamming bounds on codeword cardinality as well as the corresponding bounds on minimum distance of a code are derived. We also derive formulas for bounds on the code rate and its scaling law, which are key to establish existence results of unitary codes. In addition, a closed-form upper bound on diversity sum of unitary space-time codes is obtained. These simple-to-compute analytical results capture the behavior of the bounds reasonably accurately, which also lead to useful insights and properties of the bounds.

The rest of the paper is organized as follows. In Section~\ref{sec:problem} we formulate the problem considered in this paper, where we first define volume of a metric ball in unitary group before stating the coding bounds of interests. Section~\ref{sec:vol} is devoted to the derivation of exact and asymptotic volume formulas. In Section~\ref{sec:app} the derived analytical results are utilized in the study of coding-theoretic bounds of unitary codes. We conclude the main findings of this paper in Section~\ref{sec:con}. Proofs of some the technical results are provided in the Appendices.

\section{Problem Statement}\label{sec:problem}

\subsection{Volume of a Metric Ball}\label{subsec:DEFvol}
Consider a metric ball around the identity element $\mathbf{I}_{n}$ in the $n$-dimensional unitary group $U(n)$ with Euclidean distance (chordal distance) $r$,
\begin{equation}\label{eq:ball}
B(r)=\left\{\mathbf{U}\in U(n)~\big|~||\mathbf{U}-\mathbf{I}_{n}||_{\text{F}}\leq r\right\},
\end{equation}
where $||\cdot||_{\text{F}}$ is the Frobenius norm topological metric. We consider the invariant Haar measure $\mu$, defining a uniform distribution on $U(n)$. For any measurable set $\mathcal{S}\subset U(n)$ and any $\mathbf{U}\in U(n)$, the Haar measure satisfies $\mu\left(\mathbf{U}\mathcal{S}\right)=\mu\left(\mathcal{S}\right)$. Due to the homogeneity of $U(n)$, the characteristics of the ball~(\ref{eq:ball}) centered at any group element, say $\mathbf{A}$, would be the same.

The eigenvalue decomposition of $\mathbf{U}$ takes the form
\begin{equation}\label{eq:EVD}
\mathbf{U}=\mathbf{S}^{-1}\mathbf{E}\mathbf{S},
\end{equation}
where $\mathbf{S}\in U(n)$ and the diagonal entries of $\mathbf{E}$ are $n$ complex numbers $\e^{\imath\theta_{i}}$ on the unit circle. More precisely, $\mathbf{S}$ takes values in a flag manifold of equivalence classes of $U(n)$ modulo diagonal matrices of the form $\mathbf{E}$. The joint density of the angles $\theta_{i}$ is given by~\cite{Mehta}
\begin{equation}\label{eq:j}
p\left(\theta_{1},\dots,\theta_{n}\right)=\frac{1}{c}\prod_{1\leq j<k\leq n}\left|\e^{\imath\theta_{j}}-\e^{\imath\theta_{k}}\right|^{2},
\end{equation}
where $-\pi\leq\theta_{i}\leq\pi$, $i=1,\dots,n$ and the constant $c=n!(2\pi)^{n}$. The product $\prod_{1\leq j<k\leq n}\left|\e^{\imath\theta_{j}}-\e^{\imath\theta_{k}}\right|^{2}$ is the Jacobian of the transform~(\ref{eq:EVD}) of Haar measure $(\!\dd\mathbf{U})$ to the measure $(\!\dd\mathbf{S})\prod_{j=1}^{n}\dd\theta_{j}$. In the random matrix theory literature, the joint density~(\ref{eq:j}) is referred to as circular unitary ensemble~\cite{1962Dyson} and $c$ is the corresponding partition function.

The condition on the distance measure $||\mathbf{U}-\mathbf{I}_{n}||_{\text{F}}\leq r$ in~(\ref{eq:ball}) is equivalent to $\sum_{i=1}^{n}\sin^{2}\left(\theta_{i}/2\right)\leq r^{2}/4$. Thus, the (normalized) volume of the metric ball~(\ref{eq:ball}) equals the following $n$-dimensional integral~\cite{2006Han}
\begin{equation}\label{eq:v}
\mu\left(B\left(r\right)\right)=\int\dots\int_{\substack{-\pi\leq\theta_{i}\leq\pi, \\ \sum_{i=1}^{n}\sin^{2}\frac{\theta_{i}}{2}\leq\frac{ r^{2}}{4}}}p\left(\theta_{1},\dots,\theta_{n}\right)\prod_{j=1}^{n}\dd\theta_{j},
\end{equation}
where $0\leq r\leq2\sqrt{n}$. For the maximal distance $r=2\sqrt{n}$, the restriction $\sum_{i=1}^{n}\sin^{2}\left(\theta_{i}/2\right)\leq r^{2}/4$ becomes irrelevant and $\mu(B(2\sqrt{n}))=1$ by the definition~(\ref{eq:j}). For this reason, we also refer~(\ref{eq:v}) as a restricted partition function of the circular unitary ensemble. Finally, we note that the volume of metric balls with other distance measures such as the geodesic distance could be similarly obtained by the analytical framework developed in Section~\ref{sec:vol}.

\subsection{Coding-theoretic Bounds}\label{subsec:SPB}
A fundamental problem in coding theory is to study the maximum size of codewords as a function of the minimum distance of a code. A unitary code $\mathcal{C}$ with cardinality $|\mathcal{C}|$,
\begin{equation}\label{eq:C}
\mathcal{C}=\left\{\mathbf{U}_{1},\mathbf{U}_{2},\dots,\mathbf{U}_{|\mathcal{C}|}\right\}\subset U(n)
\end{equation}
is a finite subset of unitary group $U(n)$. With a slight abuse of notation, we define
\begin{equation}\label{eq:MDdef}
r=\min\bigg\{||\mathbf{U}_{i}-\mathbf{U}_{j}||_{\text{F}}~\big|~\mathbf{U}_{i}, \mathbf{U}_{j}\in\mathcal{C}, i\neq j\bigg\}
\end{equation}
as the minimum distance between distinct codewords in $U(n)$. Bounds on codeword cardinality in unitary group rely on the normalized volume of the metric ball $\mu\left(B\left(r\right)\right)$ formulated in~(\ref{eq:v}). In particular, the Gilbert-Varshamov lower bound and the Hamming upper bound on the cardinality $|\mathcal{C}|$ are related to volumes of metric ball as~\cite{2005Henkel,2002Barg}
\begin{equation}\label{eq:bounds}
\underbrace{\frac{1}{\mu\left(B\left(r\right)\right)}}_{\text{\normalsize Gilbert-Varshamov bound}}\leq|\mathcal{C}|\leq\underbrace{\frac{1}{\mu\left(B\left(r/2\right)\right)}}_{\text{\normalsize Hamming bound}}.
\end{equation}
The Gilbert-Varshamov bound and the Hamming bound are referred to as a sphere covering bound and a sphere packing bound, respectively. The principles of the two coding bounds are different. The Gilbert-Varshamov bound~(\ref{eq:bounds}) is based on a greedy but natural approach to construct a unitary code: start with any codeword and keep on adding codewords that have distance at least $r$ from all codewords already included until the whole space is covered. Such an algorithm will terminate at $|\mathcal{C}|\mu\left(B\left(r\right)\right)\geq1$, where $\left(|\mathcal{C}|-1\right)\mu\left(B\left(r\right)\right)<1$. The argument $r/2$ in the Hamming bound is related to the error correcting capability of the code, where errors made within `Hamming sphere' of radius $r/2$ can be corrected. This generalizes the concept of Hamming distance of q-ary block codes to unitary codes. By taking the union of disjoint metric balls each with $r/2$ packing radius, we arrive at the Hamming bound~(\ref{eq:bounds}).

Another important quantity in coding theory, especially in the limit $n\to\infty$, is rate of the code~\cite{2002Barg,2005Henkel}
\begin{equation}\label{eq:ratedef}
R=\frac{1}{n}\log_{2}|\mathcal{C}|.
\end{equation}
The bounds on cardinality~(\ref{eq:bounds}) immediately lead to bounds on code rate
\begin{equation}\label{eq:rate}
\frac{1}{n}\log_{2}\left(\frac{1}{\mu\left(B\left(r\right)\right)}\right)\leq R\leq\frac{1}{n}\log_{2}\left(\frac{1}{\mu\left(B\left(r/2\right)\right)}\right).
\end{equation}
Namely, for any $n$ there exists a code in $U(n)$ with a minimum distance $r$ and code rate as bounded in~(\ref{eq:rate}). The corresponding bounds on minimum distance $r$ as a function of code rate $R$ is obtained as
\begin{equation}\label{eq:mindist}
\mu^{-1}\left(2^{-nR}\right)\leq r\leq 2\mu^{-1}\left(2^{-nR}\right),
\end{equation}
where $\mu^{-1}(\cdot)$ denotes inverse function of the volume~(\ref{eq:v}). Note that besides the Frobenius norm other metrics such as the spectral norm can be equally considered in the definition of minimum distance~(\ref{eq:MDdef}), and the corresponding bounds in~(\ref{eq:bounds}),~(\ref{eq:rate}) and~(\ref{eq:mindist}) are still valid.

Finally, we define the diversity sum of a unitary code~\cite{2002Liang,2006bHan,2006Han}, which is closely related to the minimum distance~(\ref{eq:MDdef}), as
\begin{equation}\label{eq:diverdef}
\Sigma=\frac{r}{2\sqrt{n}}.
\end{equation}
The diversity sum~(\ref{eq:diverdef}) is an important performance measure for unitary space-time codes, where a code with large diversity sum tends to perform well at the most critical regime of low signal-to-noise ratio~\cite{2006bHan,2006Han}. Therefore, it is interesting to know the largest possible value of diversity sum for a given dimension $n$ and cardinality $|\mathcal{C}|$. A useful Hamming-type upper bound to $\Sigma$ is given by~\cite[Eq.~(B1)]{2006Han}
\begin{equation}\label{eq:diversity}
\Sigma\leq\sqrt{\frac{\left(\mu^{-1}\left(1/|\mathcal{C}|\right)\right)^{2}}{n}-\frac{\left(\mu^{-1}\left(1/|\mathcal{C}|\right)\right)^{4}}{4n^{2}}},
\end{equation}
which is tight for large $|\mathcal{C}|$. Note that the other diversity sum bounds~(B2) and~(B3) proposed in~\cite{2006Han} as well as the linear programming bounds~\cite{2010Creignou} can be also analyzed by the framework developed in Section~\ref{sec:vol}. Finally, from the diversity sum bound~(\ref{eq:diversity}) and the relation~(\ref{eq:diverdef}), we reach another upper bound on minimum distance
\begin{equation}\label{eq:mindist2}
r\leq\sqrt{4\left(\mu^{-1}\left(2^{-nR}\right)\right)^{2}-\frac{\left(\mu^{-1}\left(2^{-nR}\right)\right)^{4}}{n}}.
\end{equation}
As discussed in~\cite{2002Barg}, this modified bound is tighter than the minimum distance upper bound ~(\ref{eq:mindist}).

\section{Volume Calculations}\label{sec:vol}
As we have seen, the performance of coding-theoretic bounds~(\ref{eq:bounds}),~(\ref{eq:rate}),~(\ref{eq:mindist}),~(\ref{eq:diversity}) and~(\ref{eq:mindist2}) all rely on an accurate estimate to volume of the corresponding metric ball. In the following, we derive an exact~(\ref{eq:R1}) as well as an asymptotic~(\ref{eq:R2}) volume formulas using diverse recipes from random matrix theory. Readers who are more interested in the applications to coding theory may skip the technical details in this section.

\subsection{Exact Volume}\label{subsec:evol}
We start by rewriting the $n$-dimensional integral~(\ref{eq:v}), with the help of a Dirac delta function $\delta(\cdot)$, as
\begin{equation}\label{eq:vde}
\mu\left(B\left(r\right)\right)=\int_{0}^{\frac{r^{2}}{4}}\int\dots\int_{-\pi\leq\theta_{i}\leq\pi}\delta\left(t-\sum_{i=1}^{n}\sin^{2}\frac{\theta_{i}}{2}\right)p\left(\theta_{1},\dots,\theta_{n}\right)\prod_{j=1}^{n}\dd\theta_{j}\dd t.
\end{equation}
Inserting the Fourier representation of Dirac delta function
\begin{equation}
\delta(t-a)=\frac{1}{2\pi}\int_{-\infty}^{\infty}\e^{\imath(t-a)\nu}\dd\nu
\end{equation}
into the reformulation~(\ref{eq:vde}) and performing the integration over $t$ first, we have
\begin{equation}\label{eq:BIR}
\mu\left(B\left(r\right)\right)=\frac{1}{2\pi}\int_{-\infty}^{\infty}\frac{\imath\left(1-\e^{\imath\frac{r^{2}}{4}\nu}\right)}{\nu\e^{\imath\frac{n}{2}\nu}}D_{n}(\nu)\dd\nu,
\end{equation}
where
\begin{equation}\label{eq:MDI}
D_{n}(\nu)=\frac{1}{c}\int\dots\int_{-\pi\leq\theta_{i}\leq\pi}\prod_{1\leq j<k\leq n}\left|\e^{\imath\theta_{j}}-\e^{\imath\theta_{k}}\right|^{2}\prod_{j=1}^{n}\e^{\imath\nu\frac{\cos\theta_{j}}{2}}\dd\theta_{j}.
\end{equation}
Comparing~(\ref{eq:v}) with~(\ref{eq:BIR}) and ~(\ref{eq:MDI}), we see that the reformulation amounts to eliminating the restriction $\sum_{i=1}^{n}\sin^{2}\left(\theta_{i}/2\right)\leq r^{2}/4$ at the expense of introducing a deformation $\prod_{j=1}^{n}\e^{\imath\nu\frac{\cos\theta_{j}}{2}}$ in the integrand.

Up to now, the idea of derivation is similar to~\cite{2006Han}. To proceed further, we notice that the term
\begin{eqnarray}
\prod_{1\leq j<k\leq n}\left|\e^{\imath\theta_{j}}-\e^{\imath\theta_{k}}\right|^{2}&=&\prod_{1\leq j<k\leq n}\left(\e^{\imath\theta_{j}}-\e^{\imath\theta_{k}}\right)\prod_{1\leq j<k\leq n}\left(\e^{-\imath\theta_{j}}-\e^{-\imath\theta_{k}}\right)\\
&=&\det\left(\e^{\imath(k-1)\theta_{j}}\right)\det\left(\e^{-\imath(k-1)\theta_{j}}\right)\label{eq:MDIII}
\end{eqnarray}
is a product of two Vandermonde determinants, where $j,k=1,\dots,n$. Invoking Andr\'{e}ief's identity (see, Appendix~\ref{a:Andreief}), we have
\begin{eqnarray}
D_{n}(\nu)&=&\frac{1}{c}\int\dots\int_{-\pi\leq\theta_{i}\leq\pi}\det\left(\e^{\imath(k-1)\theta_{j}}\right)\det\left(\e^{-\imath(k-1)\theta_{j}}\right)\prod_{j=1}^{n}\e^{\imath\nu\frac{\cos\theta_{j}}{2}}\dd\theta_{j}\\
&=&\frac{n!}{c}\det\left(\int_{-\pi}^{\pi}\e^{\imath(j-k)\theta}\e^{\imath\nu\frac{\cos\theta}{2}}\dd\theta\right)\label{eq:MDII}.
\end{eqnarray}
Applying the Euler's formula, the integral inside the above $n\times n$ determinant equals the sum of the following four integrals
\begin{eqnarray}
\int_{-\pi}^{\pi}\cos\theta(j-k)\cos\left(\frac{\nu}{2}\cos\theta\right)\dd\theta&=&2\pi\cos\left(\frac{j-k}{2}\pi\right)J_{j-k}\left(\frac{\nu}{2}\right),\\
\imath\int_{-\pi}^{\pi}\cos\theta(j-k)\sin\left(\frac{\nu}{2}\cos\theta\right)\dd\theta&=&\imath2\pi\sin\left(\frac{j-k}{2}\pi\right)J_{j-k}\left(\frac{\nu}{2}\right),\\
\imath\int_{-\pi}^{\pi}\sin\theta(j-k)\cos\left(\frac{\nu}{2}\cos\theta\right)\dd\theta&=&0,\\
-\int_{-\pi}^{\pi}\sin\theta(j-k)\sin\left(\frac{\nu}{2}\cos\theta\right)\dd\theta&=&0,
\end{eqnarray}
where
\begin{equation}\label{eq:BF}
J_{k}(x)=\sum_{j=0}^{\infty}\frac{(-1)^{j}}{\Gamma(j+k+1)j!}\left(\frac{x}{2}\right)^{2j+k}
\end{equation}
denotes the Bessel function of the first kind~\cite{2007GR}. We now have
\begin{equation}\label{eq:BI}
D_{n}(\nu)=\frac{n!}{c}\det\left(2\pi\e^{\imath\frac{j-k}{2}\pi}J_{j-k}\left(\frac{\nu}{2}\right)\right)=\det\left(J_{j-k}\left(\frac{\nu}{2}\right)\right),
\end{equation}
and by the definition~(\ref{eq:BF}) the Toeplitz determinant~(\ref{eq:BI}) has the property
\begin{equation}\label{eq:TDP}
D_{n}(-\nu)=\det\left((-1)^{j-k}J_{j-k}\left(\frac{\nu}{2}\right)\right)=\det\left(J_{j-k}\left(\frac{\nu}{2}\right)\right)=D_{n}(\nu).
\end{equation}

Inserting~(\ref{eq:BI}) into~(\ref{eq:BIR}), after some manipulations, we arrive at a one-dimensional integral representation for the volume of metric ball~(\ref{eq:v}),
\begin{equation}\label{eq:R1}
\mu\left(B\left(r\right)\right)=\frac{1}{\pi}\int_{0}^{\infty}\frac{\sin\frac{n\nu}{2}+\sin\left(\frac{r^{2}}{4}-\frac{n}{2}\right)\nu}{\nu}\det\left(J_{j-k}\left(\frac{\nu}{2}\right)\right)\dd\nu.
\end{equation}

\subsection*{The Case $n=2$}
In the simplest case $n=2$, the integral~(\ref{eq:R1}) can be explicitly calculated as
\begin{equation}\label{eq:n2}
\mu\left(B\left(r\right)\right)=\frac{1}{2}+\frac{a}{2\pi}\left(G_{2,2}^{2,0}\left(a^{2}\left|\begin{array}{c}\frac{1}{2},\frac{3}{2}\\0,0\\\end{array}\right.\right)+G_{3,3}^{2,1}\left(a^{2}\left|\begin{array}{c}\frac{1}{2},\frac{1}{2},\frac{1}{2}\\0,0,-\frac{1}{2}\\\end{array}\right.\right)\right),
\end{equation}
where $a=r^{2}/4-1$ and $0\leq r\leq2\sqrt{2}$. Here
\begin{equation}
G_{p,q}^{m,n}\left(x\left|\begin{array}{c}a_{1},\ldots,a_{n},a_{n+1},\ldots,a_{p}\\b_{1},\ldots,b_{m},b_{m+1},\ldots,b_{q}\\
\end{array}\right.\right)
\end{equation}
denotes Meijer's G-function~\cite{2007GR,1990PBM} (see Appendix~\ref{a:Meijer} for a brief introduction of this special function). The proof of~(\ref{eq:n2}) can be found in Appendix~\ref{a:n2}.

\subsection{Asymptotical Volume}\label{subsec:avol}
It seems difficult to explicitly calculate the integral~(\ref{eq:R1}) for any $n$. In addition, from an information theoretical point of view, asymptotics of the bounds~(\ref{eq:bounds}),~(\ref{eq:rate}),~(\ref{eq:mindist}) and~(\ref{eq:diversity}) are more interesting. These facts motivate us to study the asymptotical volume as $n$ goes to infinity. From the representation~(\ref{eq:BIR}), this boils down to the study of the limiting behavior of $D_{n}(\nu)$, which is found to be
\begin{equation}\label{eq:A}
\lim_{n\to\infty}D_{n}(\nu)=\e^{-\frac{\nu^{2}}{16}}.
\end{equation}
In the following, we will give two distinct proofs of the above strikingly simple limiting formula. The first proof is based on matrix-variate hypergeometric functions and the associated zonal polynomials, and the second one utilizes Szeg\H{o}'s strong limit theorem on asymptotics of Toeplitz determinants. Interestingly, while the first approach seems to be more instructive and general, it appears to have received far less attention in literature.

\begin{figure}[!t]
\centering
\includegraphics[width=4.5in]{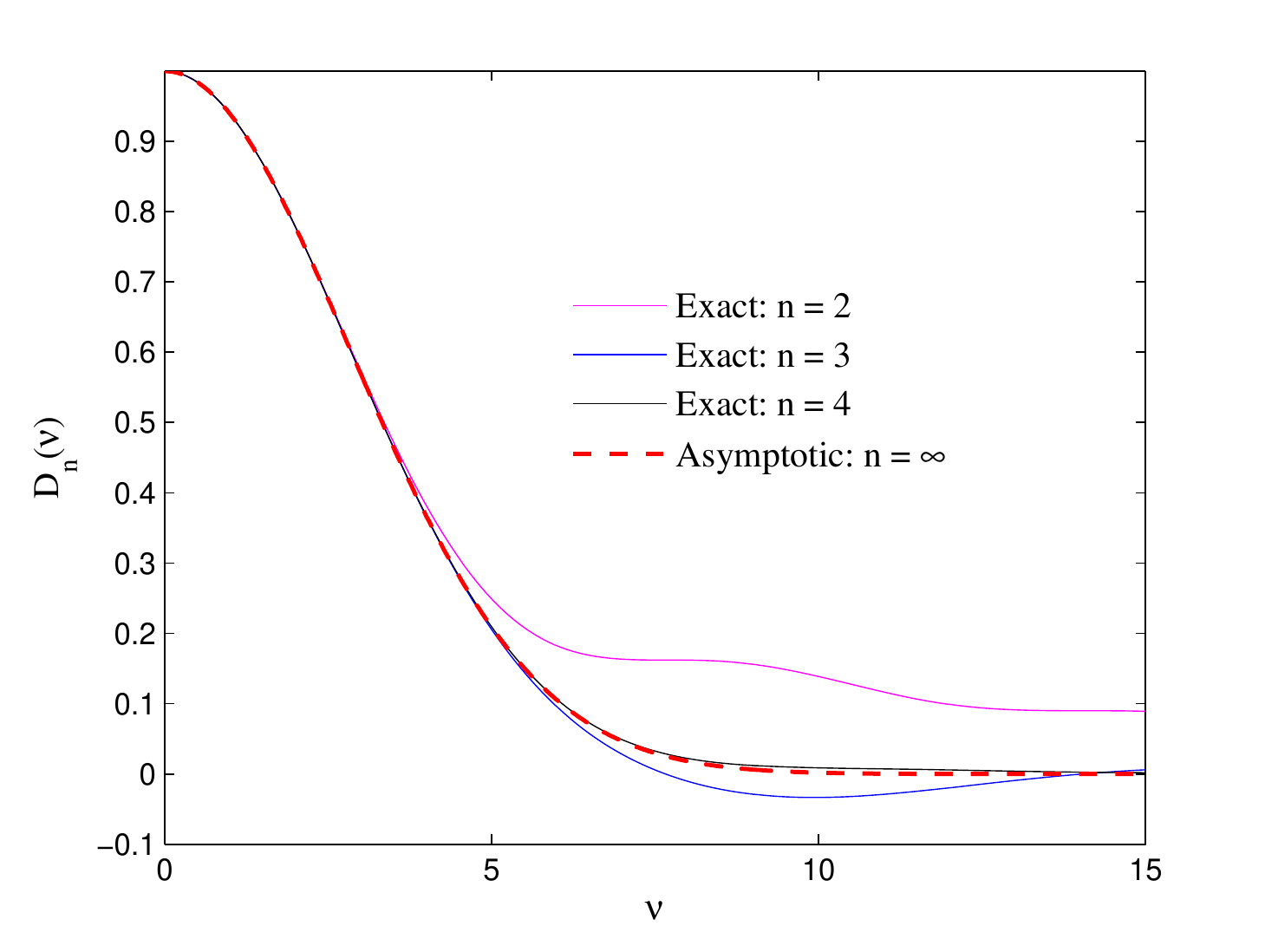}
\caption{$D_{n}(\nu)$: exact~(\ref{eq:BI}) versus asymptotic~(\ref{eq:A}).}\label{fig:p1}
\end{figure}
Whilst being simple, the convergence of $D_{n}(\nu)$ to the asymptotic limit~(\ref{eq:A}) is quite fast as observed in Figure~\ref{fig:p1}. As $n$ increases the exact curves oscillate closer to the asymptotic curve $e^{-\nu^{2}/16}$, and the difference is already indistinguishable for $n=4$. Note that due to the symmetry~(\ref{eq:TDP}), we plot only positive $\nu$ in Figure~\ref{fig:p1}.

\subsection*{Asymptotics of $D_{n}(\nu)$ via Matrix-variate Hypergeometric Functions}\label{subsec:T1}
The equality\footnote{$\Re\{\cdot\}$ denotes the real part of a complex variable.}
\begin{equation}\label{eq:LSR}
\Re\left\{\frac{1}{2}\tr\left(\mathbf{U}\right)\right\}=\frac{1}{2}\sum_{j=1}^{n}\cos\theta_{j},
\end{equation}
allows one to rewrite the $n$-dimensional integral $D_{n}(\nu)$ in~(\ref{eq:MDI}) as
\begin{equation}\label{eq:MBI}
D_{n}(\nu)=\int_{U(n)}\e^{\imath\nu\Re\left\{\frac{1}{2}\tr\left(\mathbf{U}\right)\right\}}(\!\dd\mathbf{U}),
\end{equation}
where we utilized the fact that $\prod_{1\leq j<k\leq n}\left|\e^{\imath\theta_{j}}-\e^{\imath\theta_{k}}\right|^{2}$ is the Jacobian of the transform~(\ref{eq:EVD}) from $(\!\dd\mathbf{U})$ to $(\!\dd\mathbf{S})\prod_{j=1}^{n}\dd\theta_{j}$. Here, the Haar measure ($\!\dd\mathbf{U}$) is normalized to make the total measure unity. The integral~(\ref{eq:MBI}) over the unitary group can be represented as a hypergeometric function of matrix argument~\cite[Eq.~(91)]{1964James} as
\begin{equation}
\int_{U(n)}\e^{\imath\nu\Re\left\{\frac{1}{2}\tr\left(\mathbf{U}\right)\right\}}(\!\dd\mathbf{U})=~_{0}F_{1}\left(n;-\frac{\nu^{2}}{16}\mathbf{I}_{n}\right).
\end{equation}
The function $_{0}F_{1}\left(\cdot;\cdot\right)$ is referred to as Bessel function of matrix argument~\cite{1955Herz}. It generalizes the classical univariate Bessel function expressed as an integral over the unit circle, cf. the integral inside the determinant~(\ref{eq:MDII}). The general form of matrix-variate hypergeometric function of an $n\times n$ Hermitian matrix $\mathbf{A}$ is represented as~\cite{1964James,1997Mathai}
\begin{equation}\label{eq:MHF}
_{p}F_{q}\left(a_{1},\dots,a_{p};b_{1},\dots,b_{q};\mathbf{A}\right)=\sum_{k=0}^{\infty}\sum_{\kappa}\frac{\left(a_{1}\right)_{\kappa}\cdots\left(a_{p}\right)_{\kappa}}{\left(b_{1}\right)_{\kappa}\cdots\left(b_{q}\right)_{\kappa}}\frac{C_{\kappa}(\mathbf{A})}{k!},
\end{equation}
where the sum over $\kappa$ is a sum over all partitions of integer $k$ into no more than $n$ parts, i.e. $k=\kappa_{1}+\kappa_{2}+\dots+\kappa_{n}$ with $\kappa_{1}\geq\kappa_{2}\cdots\geq\kappa_{n}\geq0$, and
\begin{equation}\label{eq:MHC}
(a)_{\kappa}=\prod_{j=1}^{n}(a-j+1)_{\kappa_{j}}=\prod_{j=1}^{n}\frac{\left(\kappa_{j}+a-j\right)!}{\left(a-j\right)!}
\end{equation}
is the multivariate hypergeometric coefficient~\cite[Eq.~(84)]{1964James}. In~(\ref{eq:MHF}), $C_{\kappa}(\mathbf{A})$ is a zonal polynomial~\cite{1964James,1997Mathai}, which is a homogenous symmetric polynomial of degree $k$ in the $n$ eigenvalues of $\mathbf{A}$. Denoting the $j$-th eigenvalue of $\mathbf{A}$ by $a_{j}$, the zonal polynomial can be represented as~\cite[Eq.~(85)]{1964James}
\begin{equation}
C_{\kappa}(\mathbf{A})=\chi_{\kappa}(1)\chi_{\kappa}(\mathbf{A}),
\end{equation}
where
\begin{equation}
\chi_{\kappa}(1)=\frac{k!\prod_{1\leq i<j\leq n}(\kappa_{i}-\kappa_{j}-i+j)}{\prod_{j=1}^{n}(\kappa_{j}+n-j)!}
\end{equation}
and
\begin{equation}\label{eq:Schur}
\chi_{\kappa}(\mathbf{A})=\frac{\det\left(a_{i}^{\kappa_{j}+n-j}\right)}{\det\left(a_{i}^{n-j}\right)}
\end{equation}
is a Schur polynomial. Schur polynomials form a basis in the space of homogeneous symmetric polynomials in $n$ variables of degree $k$ for all $k\leq n$. In particular, it holds~\cite[Eq.~(17)]{1964James}
\begin{equation}\label{eq:TF}
\tr^{k}(\mathbf{A})=\sum_{\kappa}C_{\kappa}(\mathbf{A}).
\end{equation}
In terms of the above notations, we can write $D_{n}(\nu)$ as
\begin{eqnarray}
D_{n}(\nu)&=&~_{0}F_{1}\left(n;-\frac{\nu^{2}}{16}\mathbf{I}_{n}\right)\\
&=&\sum_{k=0}^{\infty}\sum_{\kappa}\frac{1}{(n)_{\kappa}}\frac{C_{\kappa}\left(-\frac{\nu^{2}}{16}\mathbf{I}_{n}\right)}{k!} \\
&=&\sum_{k=0}^{\infty}\frac{\left(-\frac{\nu^{2}}{16}\right)^{k}}{k!}\sum_{\kappa}\frac{C_{\kappa}(\mathbf{I}_{n})}{(n)_{\kappa}}\label{eq:DnHS},
\end{eqnarray}
where the last equality is established by~(\ref{eq:Schur}). Since the leading order term in $(n-j+1)_{\kappa_{j}}$ equals $n^{\kappa_{j}}$, by the definition~(\ref{eq:MHC}), for large $n$ we have
\begin{equation}
(n)_{\kappa}=\prod_{j=1}^{n}(n-j+1)_{\kappa_{j}}\sim n^{\kappa_{1}+\dots+\kappa_{n}}=n^{k}.
\end{equation}
Using~(\ref{eq:TF}), the sum in~(\ref{eq:DnHS}) for large $n$ becomes
\begin{equation}\label{eq:FP}
\sum_{\kappa}\frac{C_{\kappa}(\mathbf{I}_{n})}{(n)_{\kappa}}\overset{n\to\infty}{=}\frac{1}{n^{k}}\sum_{\kappa}C_{\kappa}(\mathbf{I}_{n})=\frac{1}{n^{k}}\tr^{k}(\mathbf{I}_{n})=1,
\end{equation}
and we arrive at the claimed result~(\ref{eq:A}),
\begin{equation}
\lim_{n\to\infty}D_{n}(\nu)=\sum_{k=0}^{\infty}\frac{\left(-\frac{\nu^{2}}{16}\right)^{k}}{k!}=\e^{-\frac{\nu^{2}}{16}}.
\end{equation}
For a detailed account of the asymptotics of matrix variate hypergeometric functions, we refer to~\cite{2011Richards}.

We provide here an interpretation of the obtained result. By the identity~(\ref{eq:LSR}), $D_{n}(\nu)$ in~(\ref{eq:MBI}) can be understood as the characteristic function of the random variable
\begin{equation}\label{eq:X}
X=\sum_{j=1}^{n}\frac{\cos\theta_{j}}{2}.
\end{equation}
For any function $g\left(\cdot\right)$, random variables of the form $\sum_{j=1}^{n}g\left(\theta_{j}\right)$ are called linear (spectral) statistics in random matrix theory. Comparing the limiting expression of $D_{n}(\nu)$ in~(\ref{eq:A}) with the characteristic function of a Gaussian random variable $\mathcal{N}\left(\mu,\sigma^{2}\right)$,
\begin{equation}
\e^{\imath\mu\nu-\frac{1}{2}\sigma^{2}\nu^{2}},
\end{equation}
one observes that the linear statistics~(\ref{eq:X}) follows a Gaussian distribution $X\sim\mathcal{N}\left(\mu,\sigma^{2}\right)$ with mean $\mu=0$ and variance $\sigma^{2}=1/8$ as $n\to\infty$. Thus, the limiting formula~(\ref{eq:A}) can be interpreted as a central limit theorem of the linear statistics~(\ref{eq:X}). Note that this central limit theorem could be also established by a group-theoretic approach~\cite{1994Diaconis}, which did neither use matrix-variate hypergeometric functions nor zonal polynomials.

Before going to the second proof, we examine the reason behind the fast convergence of $D_{n}(\nu)$ to $\e^{-\frac{\nu^{2}}{16}}$ as $n\to\infty$. For $\mathbf{A}=\mathbf{I}_{n}$, by repeated use of L' H\^{o}pital's rule, the Schur function~(\ref{eq:Schur}) is simplified to~\cite{1970Khatri}
\begin{equation}\label{eq:cha}
\chi_{\kappa}\left(\mathbf{I}_{n}\right)=\frac{\prod_{1\leq i<j\leq n}(\kappa_{i}-\kappa_{j}-i+j)}{\prod_{j=1}^{n}(j-1)!}.
\end{equation}
As a result, the sum in~(\ref{eq:DnHS}) equals
\begin{eqnarray}
\sum_{\kappa}\frac{C_{\kappa}(\mathbf{I}_{n})}{(n)_{\kappa}}&=&\sum_{\kappa}\chi_{\kappa}(1)\frac{\chi_{\kappa}\left(\mathbf{I}_{n}\right)}{(n)_{\kappa}}\\
&=&\sum_{\kappa}\chi_{\kappa}(1)\frac{\prod_{1\leq i<j\leq n}(\kappa_{i}-\kappa_{j}-i+j)}{\prod_{j=1}^{n}\left(\kappa_{j}+n-j\right)!}\\
&=&\frac{1}{k!}\sum_{\kappa}\chi^{2}_{\kappa}(1).
\end{eqnarray}
For $k\leq n$, by the orthogonality relation, see e.g.~\cite[Chap.~\Rmnum 1.~4]{1979Macdonald} or~\cite[p.~53]{1994Diaconis},
\begin{equation}
\sum_{\kappa}\chi_{\kappa}(1)\chi_{\kappa'}(1)=\delta_{\kappa\kappa'}k!
\end{equation}
we obtain
\begin{equation}
\sum_{\kappa}\frac{C_{\kappa}(\mathbf{I}_{n})}{(n)_{\kappa}}=1,~~~\forall k\leq n.
\end{equation}
Inserting the above into~(\ref{eq:DnHS}), we see that the first $n$ terms in the power series of $D_{n}(\nu)$ equal the corresponding first $n$ terms of $\e^{-\nu^{2}/16}$ -- the $n\to\infty$ limit of $D_{n}(\nu)$. This implies that for any $n$ the first $2n+1$ moments of the random variable~(\ref{eq:X}) are exactly the same as the limiting Gaussian random variable. Namely, even for $n=2$ the first five moments are precisely the same as its Gaussian limit. We refer to this fact as finite size mock-Gaussianity. The mock-Gaussian behavior may shed some light on the rapid convergence of $D_{n}(\nu)$ seen in Figure~\ref{fig:p1}. For $n=2$, $n=3$, and $n=4$, the formal power series of $D_{n}(\nu)$ are computed via~(\ref{eq:BI}) as
\begin{equation}
D_{2}(\nu)=\underline{1-\frac{\nu^{2}}{16}+\frac{\nu^{4}}{512}}-\frac{5\nu^{6}}{147456}+\frac{7\nu^{8}}{18874368}+\mathcal{O}\left(\nu^{10}\right),
\end{equation}
\begin{equation}
D_{3}(\nu)=\underline{1-\frac{\nu^{2}}{16}+\frac{\nu^{4}}{512}-\frac{\nu^{6}}{24576}}+\frac{23\nu^{8}}{37748736}+\mathcal{O}\left(\nu^{10}\right),
\end{equation}
and
\begin{equation}
D_{4}(\nu)=\underline{1-\frac{\nu^{2}}{16}+\frac{\nu^{4}}{512}-\frac{\nu^{6}}{24576}+\frac{\nu^{8}}{1572864}}+\mathcal{O}\left(\nu^{10}\right),
\end{equation}
respectively. Comparing the above with the power series of the limit of $D_{n}(\nu)$,
\begin{equation}\label{eq:PSA}
D_{\infty}(\nu)=\e^{-\frac{\nu^{2}}{16}}=1-\frac{\nu^{2}}{16}+\frac{\nu^{4}}{512}-\frac{\nu^{6}}{24576}+\frac{\nu^{8}}{1572864}+\mathcal{O}\left(\nu^{10}\right),
\end{equation}
one indeed notices that the first $2n+1$ powers of $\nu$ in the series of $D_{n}(\nu)$ match their infinite $n$ counterpart~(\ref{eq:PSA}). Thus, the `bulk' of $D_{n}(\nu)$ must be very close to its asymptotic limit with the difference mainly resided in the tail as seen in Figure~\ref{fig:p1}. The equality in moments up to such a high order indicates that the rate of convergence is quite fast. Indeed, it was proven in~\cite{1997Johansson} that for a class of linear statistics including~(\ref{eq:X}), the convergence rate to a Gaussian limit is $\mathcal{O}\left(n^{-cn}\right)$ for some constant $c>0$. This super-exponential rate of convergence for linear statistics is uncommon in random matrix theory.

\subsection*{Asymptotics of $D_{n}(\nu)$ via Szeg\H{o}'s Strong Limit Theorem}\label{subsec:T2}
As will be seen, the asymptotical formula~(\ref{eq:A}) of $D_{n}(\nu)$ is a direct consequence of Szeg\H{o}'s strong limit theorem~\cite{1952Szego} when properly interpreted. We first rewrite the integral~(\ref{eq:MDI}) as
\begin{equation}\label{eq:TD}
D_{n}(\nu)=\frac{1}{c}\int\dots\int_{-\pi\leq\theta_{i}\leq\pi}\prod_{1\leq j<k\leq n}\left|\e^{\imath\theta_{j}}-\e^{\imath\theta_{k}}\right|^{2}\prod_{j=1}^{n}f\left(\theta_{j}\right)\dd\theta_{j},
\end{equation}
where
\begin{equation}\label{eq:f}
f\left(\theta\right)=\e^{\imath\nu\frac{\cos\theta}{2}}.
\end{equation}
Following the same steps that led~(\ref{eq:MDI}) to~(\ref{eq:MDII}), we have
\begin{equation}\label{eq:TDFC}
D_{n}(\nu)=\det\left(f_{j-k}\right),~~j,k=1,\dots,n,
\end{equation}
where
\begin{equation}
f_{j}=\frac{1}{2\pi}\int_{-\pi}^{\pi}f\left(\theta\right)\e^{-\imath j\theta}\dd\theta,~~j=0,\pm1,\pm2,\dots,
\end{equation}
can be interpreted as the Fourier coefficient of the function~(\ref{eq:f}). Thus, $D_{n}(\nu)$ is understood as the determinant of a Toeplitz matrix $(f_{j-k})_{j,k=1}^{n}$ formed by Fourier coefficients of the function $f\left(\theta\right)$. In this form, the limiting behavior of the Toeplitz determinant~(\ref{eq:TDFC}) is known and is given by Szeg\H{o}'s strong limit theorem~\cite{1952Szego}, which is stated as follows. If $f\left(\theta\right)$ is integrable over the unit circle and the sum $\sum_{j=-\infty}^{\infty}|j||\left(\ln f\right)_{j}|^{2}$ converges, where
\begin{equation}
\left(\ln f\right)_{j}=\frac{1}{2\pi}\int_{-\pi}^{\pi}\ln f\left(\theta\right)\e^{-\imath j\theta}\dd\theta,~~j=0,\pm1,\pm2,\dots,
\end{equation}
then
\begin{equation}\label{eq:SSLT}
\ln D_{n}(\nu)=n\left(\ln f\right)_{0}+\sum_{j=1}^{\infty}j\left(\ln f\right)_{j}\left(\ln f\right)_{-j}+o(1).
\end{equation}
The Landau symbol $o(1)$ is understood as
\begin{equation}\label{eq:LS}
\lim_{n\to\infty}o(1)=0,
\end{equation}
and in particular~(\ref{eq:SSLT}) is asymptotically tight. In our case there are only two non-zero Fourier coefficients for the function $\ln f\left(\theta\right)$,
\begin{eqnarray}
\left(\ln f\right)_{j}&=&\frac{\imath\nu}{4\pi}\int_{-\pi}^{\pi}\cos\theta\e^{-\imath j\theta}\dd\theta\\
&=&\frac{\imath\nu}{4\pi}\int_{-\pi}^{\pi}\cos\theta\cos j\theta\dd\theta\\
&=&\left\{\begin{array}{l l}
\imath\nu/4 & \quad j=\pm1\\
0 & \quad j=0,\pm2,\pm3,\dots.
\end{array}\right.
\end{eqnarray}
Inserting the above into~(\ref{eq:SSLT}) we establish the claimed limiting formula~(\ref{eq:A}).

Szeg\H{o}'s study of asymptotical Toeplitz determinants~\cite{1952Szego} was motivated by problems in statistical physics in 1940's. In particular, the spin to spin correlation function of two-dimensional Ising model can be written as a Toeplitz determinant~\cite{1949Kaufman}, whose asymptotics corresponds to the thermodynamics limit of such a system. Szeg\H{o}'s original proof~\cite{1952Szego} imposed strong conditions on $f(\theta)$, which were successively weakened by several mathematicians in subsequent work. The theorem was proven in the present form, among others, in~\cite{1988Johansson}. For a concise survey on recent development of asymptotical Toeplitz determinants, we refer to~\cite{2013Deift}.

We end this subsection by pointing out a by-product of this paper. Since $D_{n}(\nu)$ also equals the determinant~(\ref{eq:BI}), a limiting formula for a Toeplitz determinant of Bessel functions~(\ref{eq:BF}) is read off as
\begin{equation}\label{eq:BP}
\lim_{n\to\infty}\det\left(\begin{array}{cccc}
J_{0}\left(x\right) & J_{-1}\left(x\right) & \cdots & J_{-n+1}\left(x\right) \\
J_{1}\left(x\right) & J_{0}\left(x\right) & \cdots & J_{-n+2}\left(x\right) \\
\vdots & \vdots & \ddots & \vdots \\
J_{n-1}\left(x\right) & J_{n-2}\left(x\right) & \cdots & J_{0}\left(x\right) \\
\end{array}\right)=e^{-\frac{x^{2}}{4}},
\end{equation}
which seems new. Without using tools from matrix-variate hypergeometric functions or Szeg\H{o}'s strong limit theorem, it seems challenging to provide an elementary proof of this interesting asymptotic identity.

\subsection*{Asymptotic Volume Formula}
We are now in a position to state the main technical result of this paper. Inserting the limiting expression~(\ref{eq:A}) into~(\ref{eq:R1}), a simple integration yields an asymptotic formula of the volume~(\ref{eq:v}),
\begin{eqnarray}
\mu\left(B\left(r\right)\right)&\approx&\frac{1}{\pi}\int_{0}^{\infty}\frac{\sin\frac{n\nu}{2}+\sin\left(\frac{r^{2}}{4}-\frac{n}{2}\right)\nu}{\nu}\e^{-\frac{\nu^{2}}{16}}\dd\nu\label{eq:R2I}\\
&=&\frac{1}{2}\text{erf}(n)-\frac{1}{2}\text{erf}\left(n-\frac{r^{2}}{2}\right),\label{eq:R2}
\end{eqnarray}
where
\begin{equation}
\text{erf}(x)=\frac{2}{\sqrt{\pi}}\int_{0}^{x}\e^{-t^{2}}\dd t
\end{equation}
is Gauss error function and~(\ref{eq:R2}) is asymptotically tight as $n\to\infty$.
\begin{figure}[!t]
\centering
\includegraphics[width=4.5in]{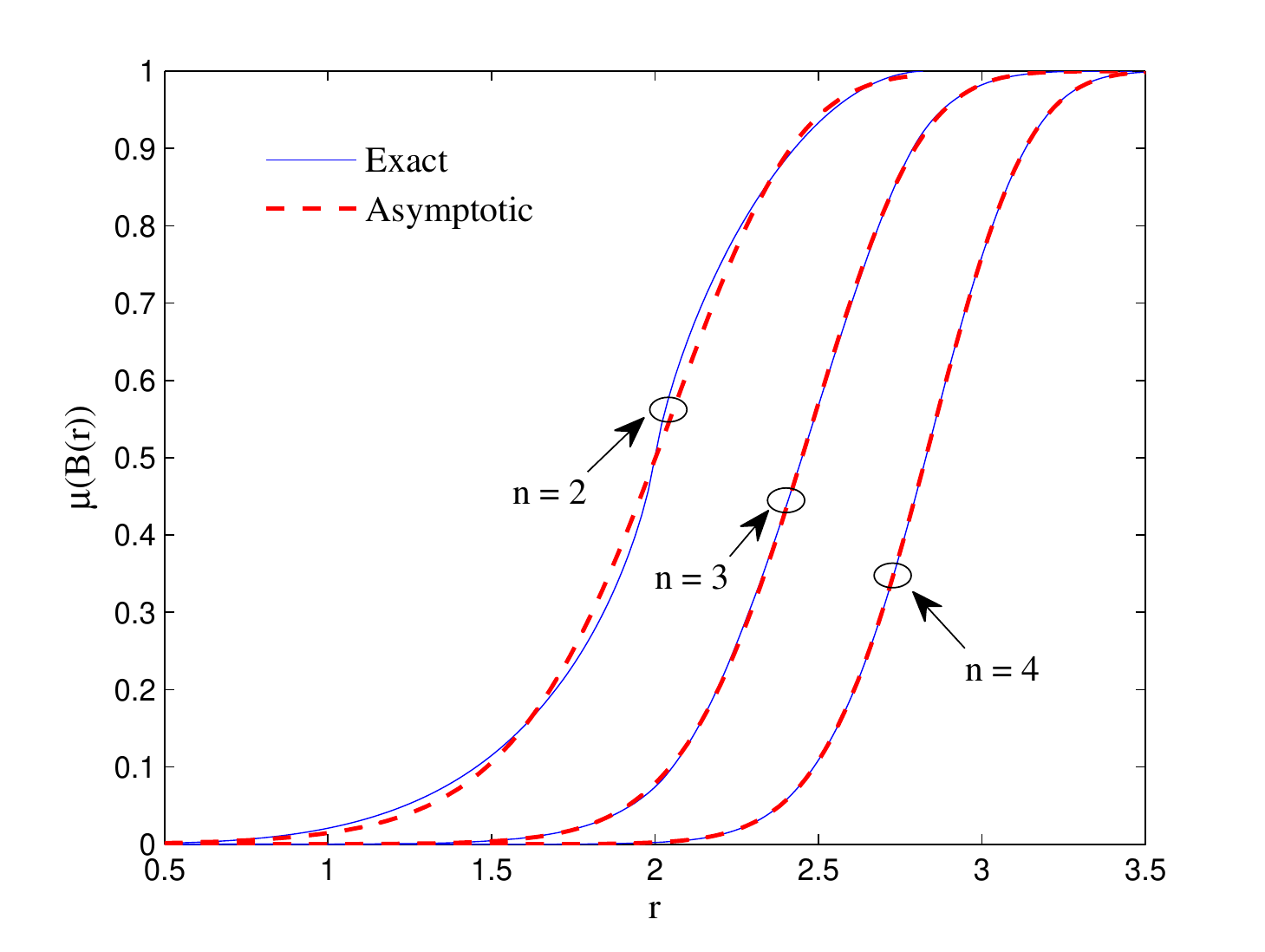}
\caption{Volumes of metric ball: exact~(\ref{eq:R1}) versus asymptotic~(\ref{eq:R2}).}\label{fig:p2}
\end{figure}
Due to the fast convergence of $D_{n}(\nu)$ as observed in Figure~\ref{fig:p1} and the rapid decreasing nature of the function $\left(\sin\frac{n\nu}{2}+\sin\left(\frac{r^{2}}{4}-\frac{n}{2}\right)\nu\right)/\nu$ in~(\ref{eq:R2I}), we expect that the asymptotic formula~(\ref{eq:R2}) approaches the exact one quite fast. Indeed, we can see from Figure~\ref{fig:p2} the approximate volume~(\ref{eq:R2}) is already very accurate for $n$ as small as three. In Figure~\ref{fig:p2}, the exact volume for $n=2$ is calculated by~(\ref{eq:n2}) and the ones for $n=3,4$ are obtained by numerical integration of~(\ref{eq:R1}).

\section{Applications to Coding Theory}\label{sec:app}
In this section, we apply the derived volume formulas to the study of coding-theoretic quantities defined in Section~\ref{subsec:SPB}. As will be shown, the asymptotic volume expression~(\ref{eq:R2}) leads to simple yet accurate analytical formulas of the bounds\footnote{Formally speaking, these are asymptotic bounds valid for $n\to\infty$. This fact may not be always mentioned in the following as the finite-size accuracy is of primary concern in practice.} on cardinality~(\ref{eq:boundsA}), code rate~(\ref{eq:rateA}) as well as its scaling law~(\ref{eq:rateAsy}), minimum distance~(\ref{eq:mindistA}),~(\ref{eq:mindistA2}) and diversity sum~(\ref{eq:diversityA}).

Inserting the limiting volume formula~(\ref{eq:R2}) into~(\ref{eq:bounds}), the Gilbert-Varshamov lower bound and the Hamming upper bound on the cardinality $|\mathcal{C}|$ of a unitary code~(\ref{eq:C}) are obtained as
\begin{equation}\label{eq:boundsA}
\frac{2}{\text{erf}(n)-\text{erf}\left(n-\frac{r^{2}}{2}\right)}\leq|\mathcal{C}|\leq\frac{2}{\text{erf}(n)-\text{erf}\left(n-\frac{r^{2}}{8}\right)}.
\end{equation}
The corresponding bounds on code rate~(\ref{eq:ratedef}) are read off as
\begin{equation}\label{eq:rateA}
\frac{1}{n}\log_{2}\left(\frac{2}{\text{erf}(n)-\text{erf}\left(n-\frac{r^{2}}{2}\right)}\right)\leq R\leq\frac{1}{n}\log_{2}\left(\frac{2}{\text{erf}(n)-\text{erf}\left(n-\frac{r^{2}}{8}\right)}\right).
\end{equation}
A notably important metric in coding theory is the scaling law of code rate, which establishes existence results of codes as dimension grows to infinity~\cite{2002Barg}. The obtained analytical bounds on code rate~(\ref{eq:rateA}) lead to elementary bounds on limiting code rate. To wit, when dimension $n$ approaches infinity with the same speed as the squared minimum distance $r^{2}$, i.e. under the regime
\begin{equation}\label{eq:regime}
n\to\infty,~r\to\infty,~\text{with}~\lambda=\frac{r^{2}}{n}~\text{fixed},
\end{equation}
by using an asymptotic expansion of error function
\begin{equation}\label{eq:erfA}
\text{erf}(x)=1-\frac{1}{\sqrt{\pi}x}\e^{-x^{2}}\left(1+\mathcal{O}\left(\frac{1}{x^{2}}\right)\right),~~~x\to\infty,
\end{equation}
and after some straightforward if tedious manipulations, it can be shown that the following limit exists
\begin{equation}\label{eq:ratelim}
\lim_{n\to\infty}\frac{1}{n^{2}}\log_{2}\left(\frac{2}{\text{erf}(n)-\text{erf}\left(n-\frac{r^{2}}{b}\right)}\right)=\frac{(\lambda-b)^{2}}{b^{2}\ln{(2)}}\mathbbm{1}_{[0,b]},
\end{equation}
where
\begin{equation}\label{eq:indic}
\mathbbm{1}_{[a,b]}=\left\{\begin{array}{l l}
1 & \quad \lambda\in[a,b]\\
0 & \quad \text{otherwise}
\end{array}\right.
\end{equation}
defines an indicator function. In~(\ref{eq:ratelim}), the limit is taken over the regime~(\ref{eq:regime}) and $\ln(\cdot)$ denotes the natural logarithm. As a direct consequence, the normalized limiting code rate $\widetilde{R}=R/n$ is bounded by
\begin{equation}\label{eq:rateAsy}
\frac{(\lambda-2)^{2}}{4\ln(2)}\mathbbm{1}_{[0,2]}\leq\widetilde{R}\leq\frac{(\lambda-8)^{2}}{64\ln(2)}\mathbbm{1}_{[0,8]}.
\end{equation}
\begin{figure}[!t]
\centering
\includegraphics[width=4.5in]{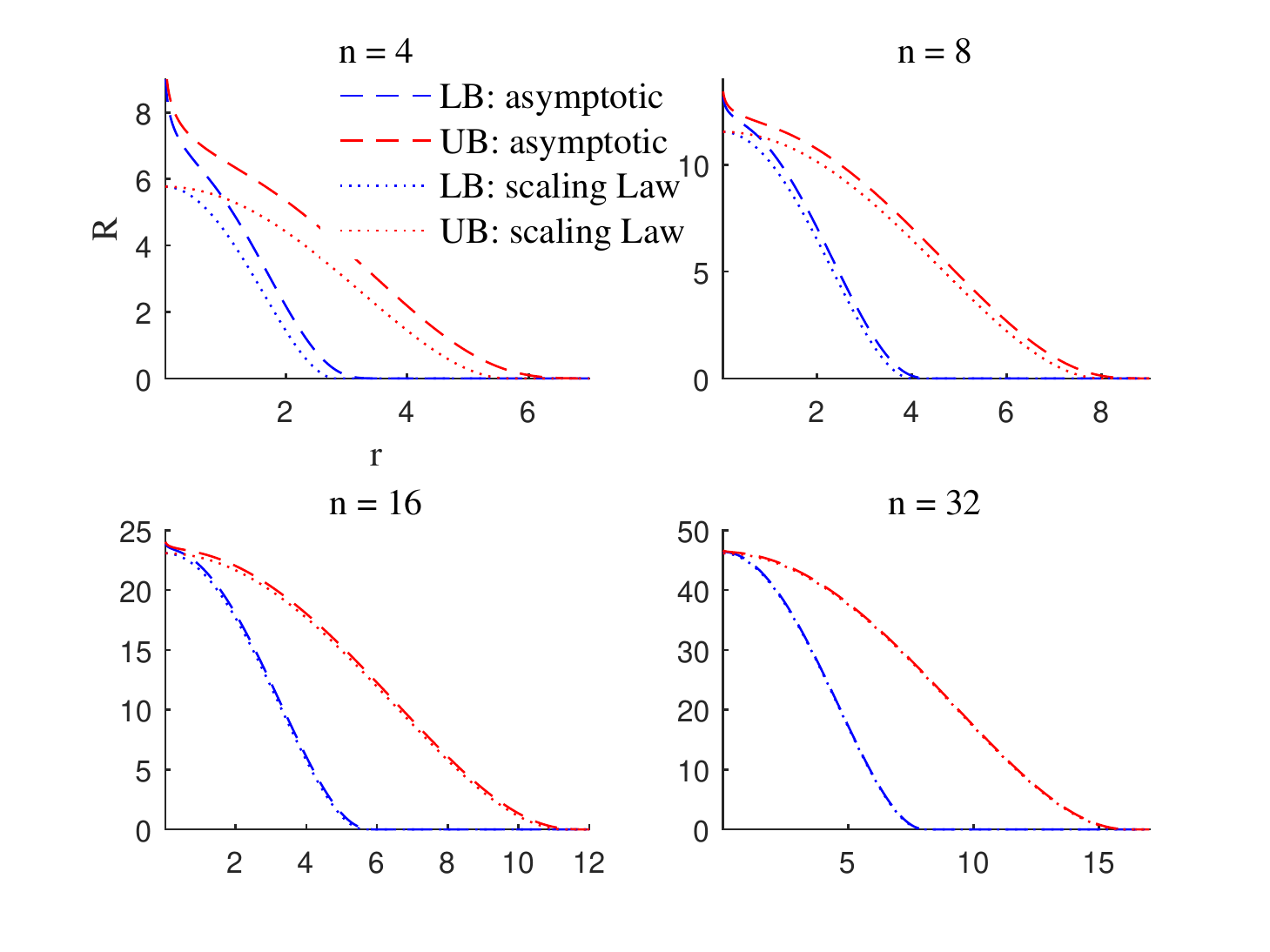}
\caption{Lower bounds (LB) and upper bounds (UB) on code rate $R$ as a function of minimum distance $r$: convergence of the asymptotic bounds~(\ref{eq:rateA}) to the bounds on scaling law~(\ref{eq:rateAsy}).}\label{fig:pr}
\end{figure}Thus, in the asymptotic regime~(\ref{eq:regime}) there exist codes in the unitary group with code rate per dimension bounded by~(\ref{eq:rateAsy}). Another interpretation of~(\ref{eq:rateAsy}) is as a scaling law of code rate bounds, where clearly the bounds scale linearly with dimension in the regime~(\ref{eq:regime}). Similar results for codes in Grassmann manifold under a different asymptotic regime can be found in~\cite[Th.~2]{2002Barg}. In Figure~\ref{fig:pr} we plot the asymptotic bounds on code rate~(\ref{eq:rateA}) and bounds on the corresponding scaling law~(\ref{eq:rateAsy}) as a function of minimum distance $r$. We see that as $n$ and $r$ increase the rate bounds~(\ref{eq:rateA}) approach the bounds on scaling law~(\ref{eq:rateAsy}), as expected. The convergence is reasonably fast as for $n=16$ the differences of the respective bounds seem indistinguishable for the range of minimum distances $r$ considered even though the scaling law bounds~(\ref{eq:rateAsy}) are formally valid only when $n\to\infty$, $r\to\infty$. It is also observed from Figure~\ref{fig:pr} that the respective lower and upper bounds merge in the high rate regime, whereas their differences increase monotonically as the code rate decreases. In fact, from the simple scaling bounds~(\ref{eq:rateAsy}) one could easily deduce that the maximum gap (at $R=0$) of minimum distances equals $\sqrt{2n}$ for any $n$.

By inverting the bounds on code rate~(\ref{eq:rateA}), the resulting bounds on minimum distance $r$ as a function of code rate $R$ is obtained as
\begin{equation}\label{eq:mindistA}
\sqrt{2n-2\text{erf}^{-1}\left(\text{erf}\left(n\right)-2^{1-nR}\right)}\leq r\leq2\sqrt{2n-2\text{erf}^{-1}\left(\text{erf}\left(n\right)-2^{1-nR}\right)},
\end{equation}
where $\text{erf}^{-1}(\cdot)$ is inverse error function.

We now turn to diversity sum~(\ref{eq:diverdef}), which is relevant to performance of unitary space-time codes. By inverting the volume formula~(\ref{eq:R2}) and inserting it into~(\ref{eq:diversity}), after some manipulations we arrive at an analytical expression for the upper bound on diversity sum
\begin{equation}\label{eq:diversityA}
\Sigma\leq\frac{1}{n}\sqrt{n^{2}-\left(\text{erf}^{-1}\left(\text{erf}(n)-\frac{2}{|\mathcal{C}|}\right)\right)^{2}}.
\end{equation}
\begin{table*}[t!]
\caption{Diversity Sum Upper Bound~(\ref{eq:diversity}): Relative Error of the Approximation~(\ref{eq:diversityA})}\centering
\ra{1.4}
\begin{tabular}{@{}ccccccccc@{}}
\toprule
$|\mathcal{C}|$ & $24$ & $48$ & $64$ & $80$ & $100$ & $120$ & $128$ & $1000$\\
\hline
$n=2$ & $5.0\times10^{-2}$ & $7.3\times10^{-2}$ & $7.8\times10^{-2}$ & $7.9\times10^{-2}$ & $7.7\times10^{-2}$ & $7.3\times10^{-2}$ & $7.1\times10^{-2}$ & $1.4\times10^{-1}$\\
$n=4$ & $7.4\times10^{-5}$ & $3.4\times10^{-4}$ & $5.4\times10^{-4}$ & $6.7\times10^{-4}$ & $7.6\times10^{-4}$ & $7.9\times10^{-4}$ & $7.9\times10^{-4}$ & $2.1\times10^{-3}$\\
$n=8$ & $1.8\times10^{-8}$ & $5.4\times10^{-9}$ & $9.9\times10^{-9}$ & $2.3\times10^{-8}$ & $3.5\times10^{-8}$ & $4.2\times10^{-8}$ & $4.4\times10^{-8}$ & $1.2\times10^{-7}$\\
\bottomrule
\end{tabular}
\label{table:ds}
\end{table*}The above result makes it possible to analytically study the diversity sum bound. In particular, it can be verified that the approximative upper bound~(\ref{eq:diversityA}) is a monotonically increasing function of codeword dimension $n$ and a monotonically decreasing function of cardinality $|\mathcal{C}|$ with a closed-form relation between $n$ and $|\mathcal{C}|$ compactly captured by~(\ref{eq:diversityA}). The exact upper bound on diversity sum~(\ref{eq:diversity}) was numerically evaluated in~\cite{2006Han}, where the authors stated that ``When $n$ is large, the exact computation of $r_{0}^{E}$ \big(the inverse volume $\mu^{-1}\left(1/|\mathcal{C}|\right)$\big) is rather involved and hence it is also computationally difficult to compute the bounds''. Indeed, inverting the integral~(\ref{eq:R1}) whose integrand comprises a determinant~(\ref{eq:BI}) is numerically unstable even for a small $n$. On the other hand, the fact that the relative approximation error\footnote{For a quantity $a$ and its estimate $\tilde{a}$, the relative error is defined as absolute value of $(a-\tilde{a})/a$.} of the simple analytical formula~(\ref{eq:diversityA}) diminishes quite fast, as illustrated in Table~\ref{table:ds}, makes its usefulness more prominent. In Table~\ref{table:ds}, we follow~\cite[Table~\Rmnum 1]{2006Han} for different cardinalities $|\mathcal{C}|$ ranging between $24$ and $1000$. It is seen that the error of~(\ref{eq:diversityA}) is smaller for not-too-large $|\mathcal{C}|$ than for $|\mathcal{C}|=1000$ corresponding to the tail of the distribution~(\ref{eq:R2}), though the error at which drops rapidly as $n$ increases. Lastly, an analytical expression for the modified bound on minimum distance~(\ref{eq:mindist2}) can be similarly obtained by inverting the volume formula~(\ref{eq:R2}) as
\begin{equation}\label{eq:mindistA2}
r\leq\frac{2}{\sqrt{n}}\sqrt{n^{2}-\left(\text{erf}^{-1}\left(\text{erf}(n)-2^{1-nR}\right)\right)^{2}}.
\end{equation}
In Table~\ref{table:md} we compare minimum distance upper bound in~(\ref{eq:mindistA}) denoted by $r_{1}$ to the modified upper bound~(\ref{eq:mindistA2}) denoted by $r_{2}$, where the values in each bracket $(\cdot,\cdot)$ are computed from $r_{1}$ and $r_{2}$, respectively. As expected, the modified upper bound $r_{2}$ is tighter than the upper bound $r_{1}$. One may also observe that the ratio of the two bounds $r_{1}/r_{2}$ grows with dimension $n$, which seems to approach to a constant for large $n$. It turns out that for a fixed code rate $R$,
\begin{equation}\label{eq:mdlim}
\lim_{n\to\infty}\frac{r_{1}}{r_{2}}=\sqrt{2}\lim_{n\to\infty}\frac{\sqrt{n}}{\sqrt{n+\text{erf}^{-1}\left(\text{erf}(n)-2^{1-nR}\right)}}=\sqrt{2},
\end{equation}
where the last equality is established by the asymptotic expansion~(\ref{eq:erfA}) as well as a corresponding expansion of inverse error function
\begin{equation}\label{eq:ierfA}
\text{erf}^{-1}(x)=\frac{1}{\sqrt{2}}\sqrt{\ln\left(\frac{2}{\pi(x-1)^{2}}\right)-\ln\left(\ln\left(\frac{2}{\pi(x-1)^{2}}\right)\right)}~,~~~x\to1.
\end{equation}
\begin{table*}[t!]
\caption{Comparisons of Minimum Distance Upper Bounds~(\ref{eq:mindistA}) and~(\ref{eq:mindistA2})}\centering
\ra{1.4}
\begin{tabular}{@{}lcccc@{}}
\toprule
 & $n=2$ & $n=4$ & $n=8$ & $n=16$ \\
\hline
$R=0.1$ & $(4.738,~2.588)$ & $(5.996,~3.969)$ & $(8.066,~5.656)$ & $(11.203,~7.998)$ \\
$R=0.5$ & $(4.004,~2.828)$ & $(5.309,~3.971)$ & $(7.438,~5.605)$ & $(10.628,~7.945)$ \\
$R=1$ & $(3.497,~2.749)$ & $(4.829,~3.850)$ & $(6.997,~5.498)$ & $(10.218,~7.863)$ \\
$R=5$ & $(0.802,~0.794)$ & $(2.251,~2.160)$ & $(4.912,~4.425)$ & $(8.379,~7.138)$ \\
$R=10$ & $(0.027,~0.027)$ & $(0.011,~0.011)$ & $(2.490,~2.429)$ & $(6.718,~6.097)$ \\
\bottomrule
\end{tabular}
\label{table:md}
\end{table*}Namely, the ratio of bounds tends to a unique constant as $n\to\infty$ for any code rate $R$. Though the upper bound $r_{1}$ is at most $\sqrt{2}$ times looser that the modified bound $r_{2}$, their difference is shrinking in the high rate regime as seen from Table~\ref{table:md}. We also see that for a sufficient large $n$ the upper and lower\footnote{Recall that the lower bound in~(\ref{eq:mindistA}) is always half of the corresponding upper bound.} bounds on minimum distance increase\footnote{Note that these bounds are not necessarily monotonic increasing functions of $n$ especially in the high rate regime, see e.g. the case $R=10$ in Table~\ref{table:md}.} with $n$. This advocates design of large dimensional space-time codes, where there potentially exists codes with improved error performance, i.e. with a larger minimum distance. A similar conclusion was drawn based on the tabulated minimum distance bounds in~\cite[p.~3448]{2005Henkel}, where, however, some entries of their tables can not be computed due to the small-ball assumption made in~\cite{2005Henkel}.

\section{Conclusion}\label{sec:con}
In this paper, we derive a limiting volume formula of metric balls in unitary group based on an exact integral representation. This simple-to-evaluate asymptotic result is obtained by means of random matrix theory. The fast convergence of the derived formula is due to the identified mock-Gaussian behavior. The proposed volume formula enables analytical characterization of coding-theoretic bounds of unitary codes. Closed-form expressions for bounds on cardinality, code rate including its scaling law, minimum distance and diversity sum have been derived. These analytical formulas help gain crucial insights into the behavior of the bounds. In particular, the largest possible minimum distance gaps between the lower and upper bounds as well as between the upper bound and a modified one are quantified. It is also found that the code rate bounds obey linear scaling laws in some asymptotic regime of interest. Possible future work includes volume computations in other compact classical groups such as orthogonal, symplectic groups, and in the Stiefel manifold.

\section*{Acknowledgment}
L. Wei is supported by the Academy of Finland (Grant 251170) and Nokia Foundation. R.-A. Pitaval is supported by the Academy of Finland (Grants 276031, 282938, 283262) and the European Science Foundation under the COST Action IC1104. J. Corander is supported by the Academy of Finland (Grant 251170). O. Tirkkonen is supported by the Academy of Finland (Grant 284725).

\appendices
\section{Andr\'{e}ief integral~\cite{1883Andreief}}\label{a:Andreief}
For two $n\times n$ matrices $\mathbf{A}(\mathbf{x})$ and $\mathbf{B}(\mathbf{x})$, with the respective $ij$-th entry being functions $A_{i}(x_{j})$ and $B_{i}(x_{j})$, and a function $f(\cdot)$ such that the integral $\int_{a}^{b}A_{i}(x)B_{j}(x)f(x)\dd x$ exists, then the following multiple integral can be evaluated as
\begin{equation}\label{eq:AI}
\int\dots\int_{\mathcal{D}}\det\big(\mathbf{A}(\mathbf{x})\big)\det\big(\mathbf{B}(\mathbf{x})\big)\prod_{i=1}^{n}f(x_{i})\mathrm{d}x_{i}=\det\left(\int_{a}^{b}A_{i}(x)B_{j}(x)f(x)\dd x\right),
\end{equation}
where $\mathcal{D}=\{a\leq x_{n}\leq\ldots\leq x_{1}\leq b\}$.

\section{Meijer's G-function}\label{a:Meijer}
The general form of Meijer's G-function~\cite{2007GR,1990PBM} is
\begin{eqnarray}
G_{p,q}^{m,n}\left(x\left|\begin{array}{c}a_{1},\ldots,a_{n},a_{n+1},\ldots,a_{p}\\b_{1},\ldots,b_{m},b_{m+1},\ldots,b_{q}\\
\end{array}\right.\right)\hspace{1in}\nonumber\\
=\frac{1}{2\pi\imath}\int_{\mathcal{L}}{\frac{\prod_{j=1}^m\Gamma\left(b_j+z\right)\prod_{j=1}^n\Gamma\left(1-a_j-z\right)}{\prod_{j=n+1}^p \Gamma\left(a_{j}+z\right)\prod_{j=m+1}^q\Gamma\left(1-b_j-z\right)}}x^{-z}\,\mathrm{d}z,\label{eq:MeijerG}
\end{eqnarray}
where the contour $\mathcal{L}$ is chosen in such a way that the poles of $\Gamma(b_j+z)$, $j=1,\dots,m$ are separated from the poles of $\Gamma\left(1-a_j-z\right)$, $j=1,\dots,n$. By definition, Meijer's G-function is Mellin transform pair of
\begin{equation}\label{eq:Mellin}
\int_{0}^{\infty}x^{z-1} G_{p,q}^{m,n}\left(x\left|\begin{array}{c}a_{1},\ldots,a_{p}\\ b_{1},\ldots,b_{q}\\\end{array}\right.\right)\mathrm{d}x=\frac{\prod_{j=1}^m\Gamma(b_j+z)\prod_{j=1}^n\Gamma\left(1-a_j-z\right)}{\prod_{j=n+1}^p\Gamma(a_{j}+z)\prod_{j=m+1}^q \Gamma\left(1-b_j-z\right)}.
\end{equation}

In the following, we list some properties and results of Meijer's G-function that will be utilized in the proof of~(\ref{eq:n2}):
\begin{itemize}
\item Shifting property
\begin{equation}\label{eq:sp}
x^{k}G_{p,q}^{m,n}\left(x\left|\begin{array}{c}a_{1},\ldots,a_{p}\\b_{1},\ldots,b_{q}\\
\end{array}\right.\right)=G_{p,q}^{m,n}\left(x\left|\begin{array}{c}a_{1}+k,\ldots,a_{p}+k\\b_{1}+k,\ldots,b_{q}+k\\
\end{array}\right.\right).
\end{equation}

\item Inverse argument formula
\begin{equation}\label{eq:ia}
G_{p,q}^{m,n}\left(\frac{1}{x}\left|\begin{array}{c}a_{1},\ldots,a_{p}\\b_{1},\ldots,b_{q}\\
\end{array}\right.\right)=G_{q,p}^{n,m}\left(x\left|\begin{array}{c}1-b_{1},\ldots,1-b_{q}\\1-a_{1},\ldots,1-a_{p}\\
\end{array}\right.\right).
\end{equation}

\item A relation with hypergeometric functions
\begin{equation}\label{eq:hf}
_{p}F_{q}\left(a_{1},\dots,a_{p};b_{1},\dots,b_{q};x\right)=\frac{\prod_{k=1}^{q}\Gamma\left(b_{k}\right)}{\prod_{k=1}^{p}\Gamma\left(a_{k}\right)}G_{p,q+1}^{1,p}\left(-x\left|\begin{array}{c}1-a_{1},\ldots,1-a_{p}\\0,1-b_{1},\ldots,1-b_{q}\\\end{array}\right.\right).
\end{equation}

\item An integration identity
\begin{equation}\label{eq:ii}
\int_{0}^{\infty}\sin cx~G_{p,q}^{m,n}\left(\alpha x^{2}\left|\begin{array}{c}a_{1},\ldots,a_{p}\\b_{1},\ldots,b_{q}\\
\end{array}\right.\right)\dd x=\frac{\sqrt{\pi}}{c}G_{p+2,q}^{m,n+1}\left(\frac{4\alpha}{c^{2}}\left|\begin{array}{c}0,a_{1},\ldots,a_{p},\frac{1}{2}\\b_{1},\ldots,b_{q}\\
\end{array}\right.\right).
\end{equation}

\item A relation between a Meijer's G-function and a hypergeometric function
\begin{eqnarray}\label{eq:r1}
&&G_{3,3}^{2,1}\left(x\left|\begin{array}{c}a_{1},a_{2},a_{3}\\b_{1},b_{2},b_{3}\\\end{array}\right.\right)=\frac{\Gamma\left(1-a_{1}+b_{1}\right)\Gamma\left(1-a_{1}+b_{2}\right)}{\left(\Gamma\left(1-a_{1}+a_{2}\right)\Gamma\left(1-a_{1}+a_{3}\right)\right)^{-1}}\frac{x^{a_{1}-1}}{\Gamma\left(a_{1}-b_{3}\right)}\times\nonumber\\
&&\!\!\!\!\!\!\!\!\!\!_{3}F_{2}\left(1-a_{1}+b_{1},1-a_{1}+b_{2},1-a_{1}+b_{3};1-a_{1}+a_{2},1-a_{1}+a_{3};\frac{1}{x}\right).
\end{eqnarray}

\item A relation between a Meijer's G-function and a hypergeometric function
\begin{eqnarray}\label{eq:r2}
G_{2,2}^{2,0}\left(x\left|\begin{array}{c}a_{1},a_{2}\\b_{1},b_{2}\\\end{array}\right.\right)=x^{b_{1}}(1-x)^{a_{1}+a_{2}-b_{1}-b_{2}-1}\Gamma\left(a_{1}+a_{2}-b_{1}-b_{2}\right)\times\nonumber\\
_{2}F_{1}\left(a_{2}-b_{2},a_{1}-b_{2};a_{1}+a_{2}-b_{1}-b_{2};1-x\right).
\end{eqnarray}
\end{itemize}

\section{Proof of (\ref{eq:n2}): the exact volume for $n=2$}\label{a:n2}
By the definition of Bessel function~(\ref{eq:BF}) it holds $J_{-n}(x)=(-1)^{n}J_{n}(x)$, thus for $n=2$,
\begin{equation}
D_{2}(\nu)=J_{0}^{2}\left(\frac{\nu}{2}\right)+J_{1}^{2}\left(\frac{\nu}{2}\right),
\end{equation}
the integral~(\ref{eq:R1}) can now be split into
\begin{equation}\label{eq:IP}
\mu\left(B\left(r\right)\right)=I_{0}\left(\frac{r^{2}}{4}-1\right)+I_{0}(1)+I_{1}\left(\frac{r^{2}}{4}-1\right)+I_{1}(1),
\end{equation}
where
\begin{equation}
I_{n}(b)=\frac{1}{\pi}\int_{0}^{\infty}\frac{\sin b\nu}{\nu}J_{n}^{2}\left(\frac{\nu}{2}\right)\dd\nu.
\end{equation}
Since~\cite[Eq.~8.442]{2007GR},
\begin{equation}
J_{n}^{2}\left(\frac{\nu}{2}\right)=\frac{\left(\nu/2\right)^{2n}}{2^{2n}\Gamma^{2}(n+1)}~_{1}F_{2}\left(n+\frac{1}{2};n+1,2n+1;-\frac{\nu^{2}}{4}\right),
\end{equation}
by the relation to Meijer's G-function~(\ref{eq:hf}), we obtain
\begin{eqnarray}
J_{0}^{2}\left(\frac{\nu}{2}\right)&=&\frac{1}{\sqrt{\pi}}G_{1,3}^{1,1}\left(\frac{\nu^{2}}{4}\left|\begin{array}{c}\frac{1}{2}\\0,0,0\\\end{array}\right.\right),\\
J_{1}^{2}\left(\frac{\nu}{2}\right)&=&\frac{\nu^{2}}{8\Gamma(3/2)}G_{1,3}^{1,1}\left(\frac{\nu^{2}}{4}\left|\begin{array}{c}-\frac{1}{2}\\0,-1,-2\\\end{array}\right.\right).
\end{eqnarray}
By the shifting property~(\ref{eq:sp}), the inverse argument formula~(\ref{eq:ia}) and the integration identity~(\ref{eq:ii}), we have
\begin{eqnarray}
I_{0}(b)&=&\frac{1}{2\pi\sqrt{\pi}}\int_{0}^{\infty}\sin b\nu~ G_{1,3}^{1,1}\left(\frac{\nu^{2}}{4}\left|\begin{array}{c}0\\-\frac{1}{2},-\frac{1}{2},-\frac{1}{2}\\\end{array}\right.\right)\dd\nu\\
&=&\frac{1}{2\pi}b^{-1}G_{3,3}^{1,2}\left(b^{-2}\left|\begin{array}{c}0,0,\frac{1}{2}\\-\frac{1}{2},-\frac{1}{2},-\frac{1}{2}\\\end{array}\right.\right)\\
&=&\frac{b}{2\pi}G_{3,3}^{2,1}\left(b^{2}\left|\begin{array}{c}\frac{1}{2},\frac{1}{2},\frac{1}{2}\\0,0,-\frac{1}{2}\\\end{array}\right.\right)\label{eq:IP1}.
\end{eqnarray}
The case $I_{0}(1)$ is obtained by using the representation~(\ref{eq:r1}) as
\begin{equation}\label{eq:IP2}
I_{0}(1)=\frac{\Gamma^{2}(1/2)}{2\pi}~_{3}F_{2}\left(\frac{1}{2},\frac{1}{2},0;1,1;1\right)=\frac{1}{2}.
\end{equation}
The integral $I_{1}(b)$ is calculated by following the procedure of $I_{0}(b)$ as
\begin{eqnarray}
I_{1}(b)&=&\frac{1}{2\pi\sqrt{\pi}}\int_{0}^{\infty}\sin b\nu~ G_{1,3}^{1,1}\left(\frac{\nu^{2}}{4}\left|\begin{array}{c}0\\\frac{1}{2},-\frac{1}{2},-\frac{3}{2}\\\end{array}\right.\right)\dd\nu\\
&=&\frac{1}{2\pi}b^{-1}G_{3,3}^{1,2}\left(b^{-2}\left|\begin{array}{c}0,0,\frac{1}{2}\\ \frac{1}{2},-\frac{1}{2},-\frac{3}{2}\\\end{array}\right.\right)\\
&=&\frac{b}{2\pi}G_{3,3}^{2,1}\left(b^{2}\left|\begin{array}{c}-\frac{1}{2},\frac{1}{2},\frac{3}{2}\\0,0,-\frac{1}{2}\\\end{array}\right.\right)\\
&=&\frac{b}{2\pi}G_{2,2}^{2,0}\left(b^{2}\left|\begin{array}{c}\frac{1}{2},\frac{3}{2}\\0,0\\\end{array}\right.\right),\label{eq:IP3}
\end{eqnarray}
where the last equality is established by the definition of Meijer's G-function~(\ref{eq:MeijerG}). The case $I_{1}(1)$ is obtained by using the representation~(\ref{eq:r2}) as
\begin{equation}\label{eq:IP4}
I_{1}(1)=0.
\end{equation}
Inserting~(\ref{eq:IP1}),~(\ref{eq:IP2}),~(\ref{eq:IP3}) and~(\ref{eq:IP4}) into~(\ref{eq:IP}), we obtain the claimed result~(\ref{eq:n2}).


\begin{thebibliography}{99}
\bibitem{2001Shokrollahi}
A. Shokrollahi, B. Hassibi, B. M. Hochwald, and W. Sweldens,
\newblock ``Representation theory for high-rate multiple-antenna code design,'' {\em IEEE Trans. Inf. Theory,} vol. 47, no. 6, pp. 2335-2367, Sept. 2001.

\bibitem{2002Liang}
X.-B. Liang and X.-G. Xia,
\newblock ``Unitary signal constellations for differential space-time modulation with two transmit antennas: Parametric codes, optimal designs and bounds,'' {\em IEEE Trans. Inf. Theory,} vol. 48, no. 8, pp. 2291-2322, Aug. 2002.

\bibitem{2006bHan}
G. Han and J. Rosenthal,
\newblock ``Geometrical and numerical design of structured unitary space-time constellations,'' {\em IEEE Trans. Inf. Theory,} vol. 52, no. 8, pp. 3722-3735, Aug. 2006.

\bibitem{2006Han}
G. Han and J. Rosenthal,
\newblock ``Unitary space-time constellation analysis: an upper bound for the diversity,'' {\em IEEE Trans. Inf. Theory,} vol. 52, no. 10, pp. 4713-4721, Oct. 2006.

\bibitem{2010Creignou}
J. Creignou and H. Diet,
\newblock ``Linear programming bounds for unitary codes,'' {\em Adv. Math. Commun.,} vol. 4, no. 3, pp. 323-344, Aug. 2010.

\bibitem{2008Dai}
W. Dai, Y. Liu, and B. Rider,
\newblock ``Quantization bounds on Grassmann manifolds and applications to MIMO communications,'' {\em IEEE Trans. Inf. Theory,} vol. 54, no. 3, pp. 1108-1123, Mar. 2008.

\bibitem{2013Krishnamachari}
R. T. Krishnamachari and M. K. Varanasi,
\newblock ``On the geometry and quantization of manifolds of positive semi-definite matrices,'' {\em IEEE Trans. Signal Process.,} vol. 61, no. 18, pp. 4587-4599, Sept. 2013.

\bibitem{2002Barg}
A. Barg and D. Nogin,
\newblock ``Bounds on packings of spheres in the Grassmann manifold,'' {\em IEEE Trans. Inf. Theory,} vol. 48, no. 9, pp. 2450-2454, Sept. 2002.

\bibitem{2005Henkel}
O. Henkel,
\newblock ``Sphere-packing bounds in the Grassmann and Stiefel manifolds,'' {\em IEEE Trans. Inf. Theory,} vol. 51, no. 10, pp. 3445-3456, Oct. 2005.

\bibitem{2008Krishnamachari}
R. T. Krishnamachari and M. K. Varanasi,
\newblock ``Volume of geodesic balls in the complex Stiefel manifold,'' in {\em Proc. Allerton Conf. on Comm., Control and Comp.,} Sept. 2008.

\bibitem{2012Pitaval}
R.-A. Pitaval and O. Tirkkonen,
\newblock ``Volume of ball and Hamming-type bounds for Stiefel manifold with Euclidean distance,'' in {\em Proc. Asilomar Conf. on Sig., Systems, and Comp.,} Nov. 2012.

\bibitem{Mehta}
M. L. Mehta,
\newblock {\em Random Matrices.} $3$rd Ed., Singapore: Elsevier, 2004.

\bibitem{1962Dyson}
F. J. Dyson,
\newblock ``Statistical theory of the energy levels of complex systems \Rmnum 1,'' {\em J. Math. Phys.,} \textbf{3}, 1962.

\bibitem{1883Andreief}
C. Andr\'{e}ief,
\newblock ``Note sur une relation les int\'{e}grales d\'{e}finies des produits des fonctions,'' M\'{e}m. de la Soc. Sci. Bordeaux 2, 1883.

\bibitem{2007GR}
I. S. Gradshteyn and I. M. Ryzhik,
\newblock {\em Table of Integrals, Series, and Products.} $7$th Ed., San Diego: Academic Press, 2007.

\bibitem{1990PBM}
A. P. Prudnikov, Yu. A. Brychkov, and O. I. Marichev,
\newblock {\em Integrals and Series. Volume 3: More special functions.} New York: Gordon and Breach Science, 1990.

\bibitem{1964James}
A. T. James,
\newblock ``Distributions of matrix variates and latent roots derived from normal samples,'' {\em Ann. Math. Statist.,} vol. 35, no. 2, pp. 475-501, 1964.

\bibitem{1955Herz}
C. S. Herz,
\newblock ``Bessel functions of matrix argument,'' {\em Ann. Math.,} vol. 61, no. 3, pp. 474-523, May 1955.

\bibitem{1997Mathai}
A. M. Mathai,
\newblock {\em Jacobians of Matrix Transformations and Functions of Matrix Arguments.} Singapore: World Scientific, 1997.

\bibitem{2011Richards}
D. St. P. Richards,
\newblock ``High-dimensional random matrices from the classical matrix groups, and generalized hypergeometric functions of matrix argument,'' {\em Symmetry,} vol. 3, pp. 600-610, 2011.

\bibitem{1994Diaconis}
P. Diaconis and M. Shahshahani,
\newblock ``On the eigenvalues of random matrices,'' {\em J. Appl. Probab.,} vol. 31, pp. 49-62, 1994.

\bibitem{1970Khatri}
C. G. Khatri,
\newblock ``On the moments of traces of two matrices in three situations for complex multivariate normal populations,'' {\em Sankhy\={a},} vol. 32, pp. 65-80, 1970.

\bibitem{1979Macdonald}
I. G. Macdonald,
\newblock {\em Symmetric functions and Hall polynomials.} Oxford: Oxford University Press, 1979.

\bibitem{1997Johansson}
K. Johansson,
\newblock ``On random matrices from the compact classical groups,'' {\em Ann. Math.,} vol. 145, no. 3, pp. 519-545, May 1997.

\bibitem{1952Szego}
G. Szeg\H{o},
\newblock ``On certain Hermitian forms associated with the Fourier series of a positive function,'' {\em Comm. S\'{e}m. Math. Univ. Lund,} 1952.

\bibitem{1949Kaufman}
B. Kaufman and L. Onsager,
\newblock ``Crystal statistics. \Rmnum 3. Short-range order in a binary Ising lattice,'' {\em Phys. Rev.,} vol. 76 , no. 8, 1244-1252, 1949.

\bibitem{1988Johansson}
K. Johansson,
\newblock ``On Szeg\H{o}'s asymptotic formula for Toeplitz determinants and generalizations,'' {\em Bull. Sci. Math. (2),} \textbf{112}, 1988.

\bibitem{2013Deift}
P. Deift, A. Its, and I. Krasovsky,
\newblock ``Toeplitz matrices and Toeplitz determinants under the impetus of the Ising model: Some history and some recent results,'' {\em Comm. Pure Appl. Math.,} vol. 66, Sept. 2013.

\end{thebibliography}
\end{document}